\documentclass{vgtc}                          % final (conference style)
\ifpdf%                                % if we use pdflatex
  \pdfoutput=1\relax                   % create PDFs from pdfLaTeX
  \usepackage{graphicx}                % allow us to embed graphics files
  \DeclareGraphicsExtensions{.pdf,.png,.jpg,.jpeg} % for pdflatex we expect .pdf, .png, or .jpg files
\else%                                 % else we use pure latex
  \usepackage{graphicx}                % allow us to embed graphics files
  \DeclareGraphicsExtensions{.eps}     % for pure latex we expect eps files
\fi%

\graphicspath{{figures/}{pictures/}{images/}{./}} % where to search for the images

\usepackage{microtype}                 % use micro-typography (slightly more compact, better to read)
\PassOptionsToPackage{warn}{textcomp}  % to address font issues with \textrightarrow
\usepackage{textcomp}                  % use better special symbols
\usepackage{mathptmx}                  % use matching math font
\usepackage{times}                     % we use Times as the main font
\usepackage{cite}                      % needed to automatically sort the references
\usepackage{tabu}                      % only used for the table example
\usepackage{booktabs}                  % only used for the table example
\usepackage[T1]{fontenc}
\usepackage{amssymb}

\onlineid{0}

\vgtccategory{Research}

\vgtcinsertpkg

\title{An Interactive Interpretability System for Breast Cancer Screening with Deep Learning}

\author{Yuzhe Lu\thanks{e-mail: yuzhe.lu@vanderbilt.edu}\\ %
        \scriptsize Vanderbilt University %
\and Adam Perer\thanks{e-mail: adamperer@cmu.edu}\\ %
     \scriptsize Carnegie Mellon University. %
}

\teaser{
  \centering
  \includegraphics[width=\linewidth]{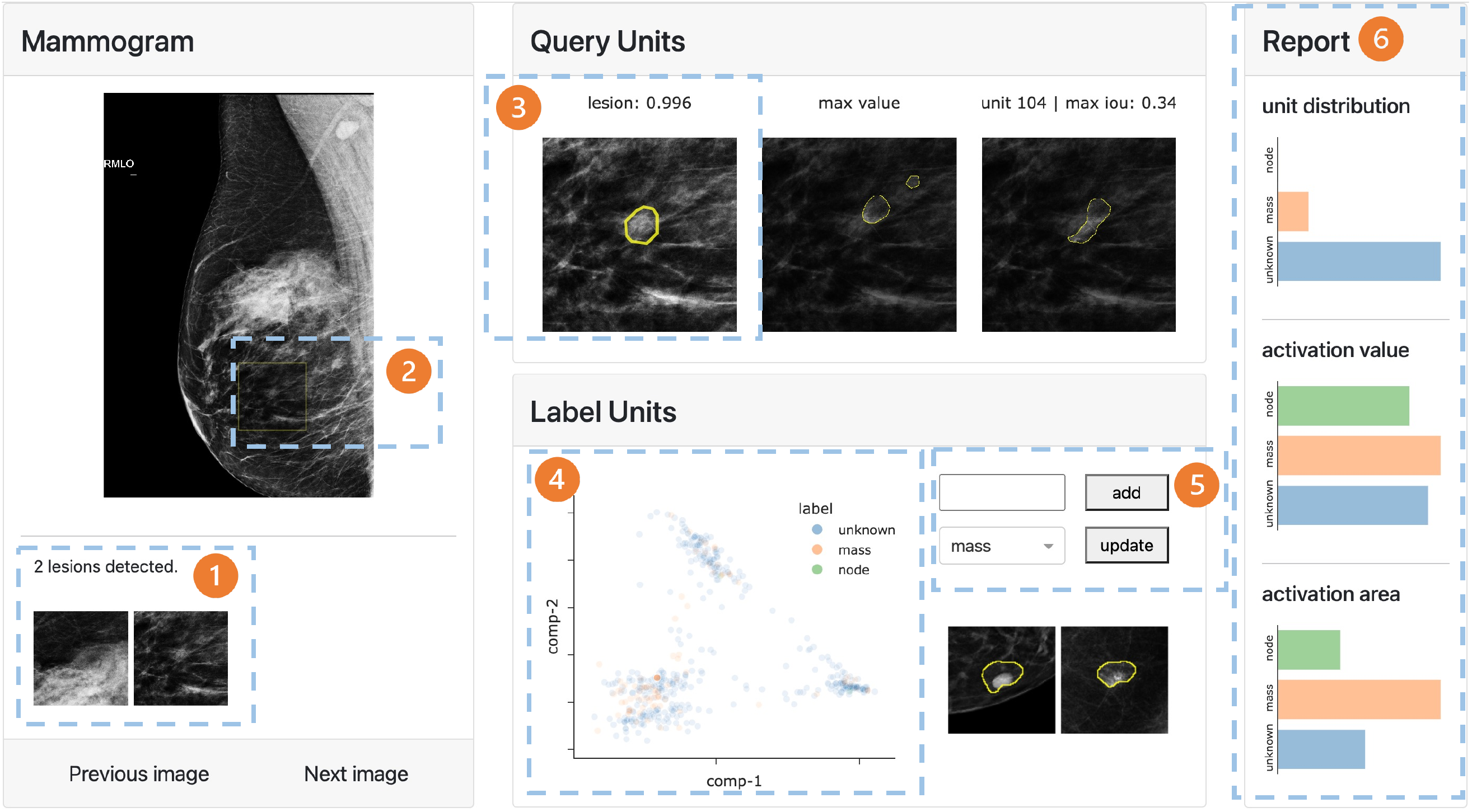}
  \caption{Overview of our interface. (1) potentially malignant patches identified by a patch-based model on the full mammogram; (2) context of a selected patch in the original mammogram; (3) place for users to query model representation with salient regions; (4) visualization of neurons based on their learned representation; (5) place for user to annotate neurons' semantic meaning; (6) explainability reports generated based on neuron annotations.}
  
  \label{fig:teaser}
}

\begin{document}


%% Abstract section.
\abstract{Deep learning methods, in particular convolutional neural networks, have emerged as a powerful tool in medical image computing tasks. While these complex models provide excellent performance, their black-box nature may hinder real-world adoption in high-stakes decision-making. In this paper, we propose an interactive system to take advantage of state-of-the-art interpretability techniques to assist radiologists with breast cancer screening. Our system integrates a deep learning model into the radiologists' workflow and provides novel interactions to promote understanding of the model's decision-making process. Moreover, we demonstrate that our system can take advantage of user interactions progressively to provide finer-grained explainability reports with little labeling overhead. Due to the generic nature of the adopted interpretability technique, our system is domain-agnostic and can be used for many different medical image computing tasks, presenting a novel perspective on how we can leverage visual analytics to transform originally static interpretability techniques to augment human decision making and promote the adoption of medical AI. 
} % end of abstract

%% ACM Computing Classification System (CCS). 
%% See <http://www.acm.org/about/class> for details.
%% We recommend the 2012 system <http://www.acm.org/about/class/class/2012>
%% For the 2012 system use the ``\CCScatTwelve'' which command takes four arguments.
%% The 1998 system <http://www.acm.org/about/class/class/2012> is still possible
%% For the 1998 system use the ``\CCScat'' which command takes four arguments.
%% In both cases the last two arguments (1998) or last three (2012) can be empty.

\CCScatlist{
  \CCScatTwelve{Human-centered computing}{Visualization}{Visualization systems and tools}{Visualization toolkits};
  \CCScatTwelve{Computing methodologies}{Machine learning}{Machine learning approaches}{Neural networks};
%   \CCScatTwelve{Applied computing}{Life and medical sciences}{Computational biology}{Imaging}
}

\firstsection{Introduction}

\maketitle

Breast cancer is the most common cancer among women worldwide and their second leading cause of death. Although early detection and treatment can improve the prognosis of a patient, screening tests have high error rates. Recently, the use of deep learning and big data has made it possible to develop high-performing models for breast cancer screening. In particular, convolutional neural networks (CNNs) have achieved remarkable performance in screening mammography \cite{agarwal2019automatic,ribli2018detecting}. Moreover, recent reader studies have shown that deep neural networks could enhance radiologists' performance in breast cancer screening \cite{wu2019deep}.

However, the deployment of CNNs in medical domain has its unique challenges. One of the key obstacles is how to effectively allow models and physicians to collaborate effectively on their complementary set of strengths. While CNNs have achieved high performance in various medical imaging domains, it is hard for physicians to understand the model's decision-making process due to its black-box nature. To tackle this challenge, both machine learning (ML) and visualization researchers have made great efforts. Many ML researchers have proposed novel methods to visualize and interpret convolutional neural networks \cite{liu2016towards, zhou2016learning, bau2020understanding, hohman2019s}. Mostly, the use of saliency maps (i.e, highlighting image features important to the model's decision) plays a central role in these methods. With this observation, a recent work \cite{boggust2022shared} proposes a system that compares saliency maps from deep neural networks to ground truth segmentations of image components to measure Human-AI alignment. 

While many ML researchers continued to polish interpretability techniques for deep learning models, few have considered these techniques being applied to medical domains in actual clinical workflows. For example, Wu et al. \cite{wu2018deepminer} proposed DeepMiner, where they applied Network Dissection \cite{bau2017network} on breast cancer screening models to uncover implicitly learned finer-grained medical concepts to improve interpretability. Although this framework provides an efficient human-in-the-loop paradigm to understand medical CNNs, it is not practical in real clinical settings as the framework requires radiologists to finish labeling neurons of a model before even using it, which adds a huge time overhead. Meanwhile, numerous visual analytics interfaces \cite{cheng2021vbridge, 8440842} have been proposed for machine learning models in healthcare applications such as electronic medical records to improve user understanding and support clinical workflow. Given the huge potential of interactive data visualizations in promoting human understanding of complex deep learning models, it's appealing to find a visual analytics solution to integrate state-of-the-art interpretability techniques into clinical workflows with deep learning components to promote Human-AI collaboration and maximize the utility of powerful medical AI models.

To this end, we proposed a novel visual analytics system to transform a generic yet powerful interpretability methodology, Network Dissection \cite{bau2017network, bau2020understanding}, and integrate it into radiologists' workflow assisted by deep learning models. Such a system may promote appropriate reliance and adoption of breast cancer screening models. The merits of our system are listed as follows:

\begin{itemize}
  \item The system empowers radiologists to interactively probe medical AI models by asking whether the model pays attention to certain features when making decisions.
  \item The system leverages interactions from the radiologist in their workflow to progressively accumulate understanding of the model to provide additional insights.
\end{itemize}

\section{Methods}

In this section, we will describe the dataset and the deep learning model used for our task, and the interpretability technique utilized in our interactive system.

\subsection{Dataset}
Our mammography dataset is from the University of Pittsburgh Medical Center (UPMC) and has been properly deidentified. Each patient in this dataset has multiple imaging views stored in DICOM format. When suspicious regions are present, a low-resolution copy with radiologist's annotations (ellipses with white boundaries) are saved. After filtering the dataset based on the DICOM header and label information, we identified 2237 patients assigned BIRADS score 0 (whose imaging contains possibly malignant findings) and 2237 patients assigned BIRADS score 1 (whose imaging contains no findings) and BIRADS score 2 (whose imaging contains benign findings) as relevant to our study.

Given these data, we aimed to build a binary classifier that differentiates mammograms of BIRADS 0 patients (possibly malignant) from those of BIRADS 1 and 2 combined (no finding or benign). As it is challenging to fit full high-resolution mammograms into a standard GPU's memory to train deep neural networks, we followed the procedures in \cite{agarwal2019automatic} to build a patch-based model. To extract patches from mammograms with BIRADS score 0, we first detected radiologists' annotation (typically a white ellipse) in the low-resolution copy and mapped the region to the corresponding high-resolution mammograms; then, we cropped square patches within the ellipse region of the high-resolution mammograms in a sliding-window fashion. We used a patch size of 512 and a step size of 256, extracting 4710 patches from 2237 BIRADS 0 patients' mammograms. To extract normal patches, we cropped $\lceil 4710 / 2237 \rceil$ patches of size 512 x 512 from all-tissue areas from mammograms of 2237 BIRADS 1 and 2 patients and uniformly sampled 4710 normal patches to create a balanced dataset.

\subsection{Model}
We used a classic convolutional neural network (CNN), VGG16\cite{simonyan2014very}, which consists of 5 blocks of $3\times3$ convolutional filters and max pooling layers, 3 fully connected layers, and a softmax activation. As the extracted patches are gray-scale images (512 x 512 x 1), we modified the VGG16 model to accept single-channel input.

To train the model, we split our dataset into train, validation, and test set with a 8:1:1 ratio. During training, we applied data augmentations (random horizontal flip, Gaussian blur) to increase sample diversity. All data are normalized to have $mean=0.5$ and $std=0.5$ before sent to the model. We used cross entropy loss as the objective function and trained the model using stochastic gradient descent with the Adam optimizer \cite{kingma2014adam}. After sweeping hyperparameters using Ray Tune \cite{liaw2018tune}, we set the batch size to $32$ and the learning rate to $1\mathrm{e}{-4}$ and trained the model for 50 epochs. The model with the lowest loss on the validation set was selected for testing. Our model achieved an area under the ROC curve (AUC) score of $0.943$ on our test set. Note that our goal is not to build the best-performing model, but rather to develop a well-performing model to prototype and experiment with our interface.

\subsection{Decode Individual Neurons}
The interpretability technique used in this work is mainly based on the Network Dissection methodology \cite{bau2017network}, which aims to quantify interpretability of individual neurons in a deep CNN. One way of determining the semantic meaning of neurons in a neural network is to look at the characteristics of images that the neuron consistently activates on. A more straightforward formulation is to simply look at the top activated images for each neuron. The process of identifying these images can be described as the following:

For a set of images $ X = \{x_{i}\}_{i=1}^{n}$ and a set of neurons $ N = \{n_{i}\}_{i=1}^{m}$. For each neuron $n_{i}$, we gather the maximum activation value $a_{i}$ on each $x_{i}$. Then, images with the topk activations will be used as the top k activated images for neuron $n_{i}$.

With top activated images, people could decide which concept a neuron captures. While deciding on concepts captured by a neuron can be done in a scalable way with segmentation models \cite{bau2017network,bau2020understanding} on natural dataset, such a task can be challenging on a medical data set, as training a segmentation model would require a more densely labeled data set, which is generally not available. Thus, the nature of medical image dataset necessitates novel strategies to decode the semantic meaning of neurons in medical AI models to assist user's understanding. 

\begin{figure*}[!t]
    \centering
    \includegraphics[width=0.8\linewidth]{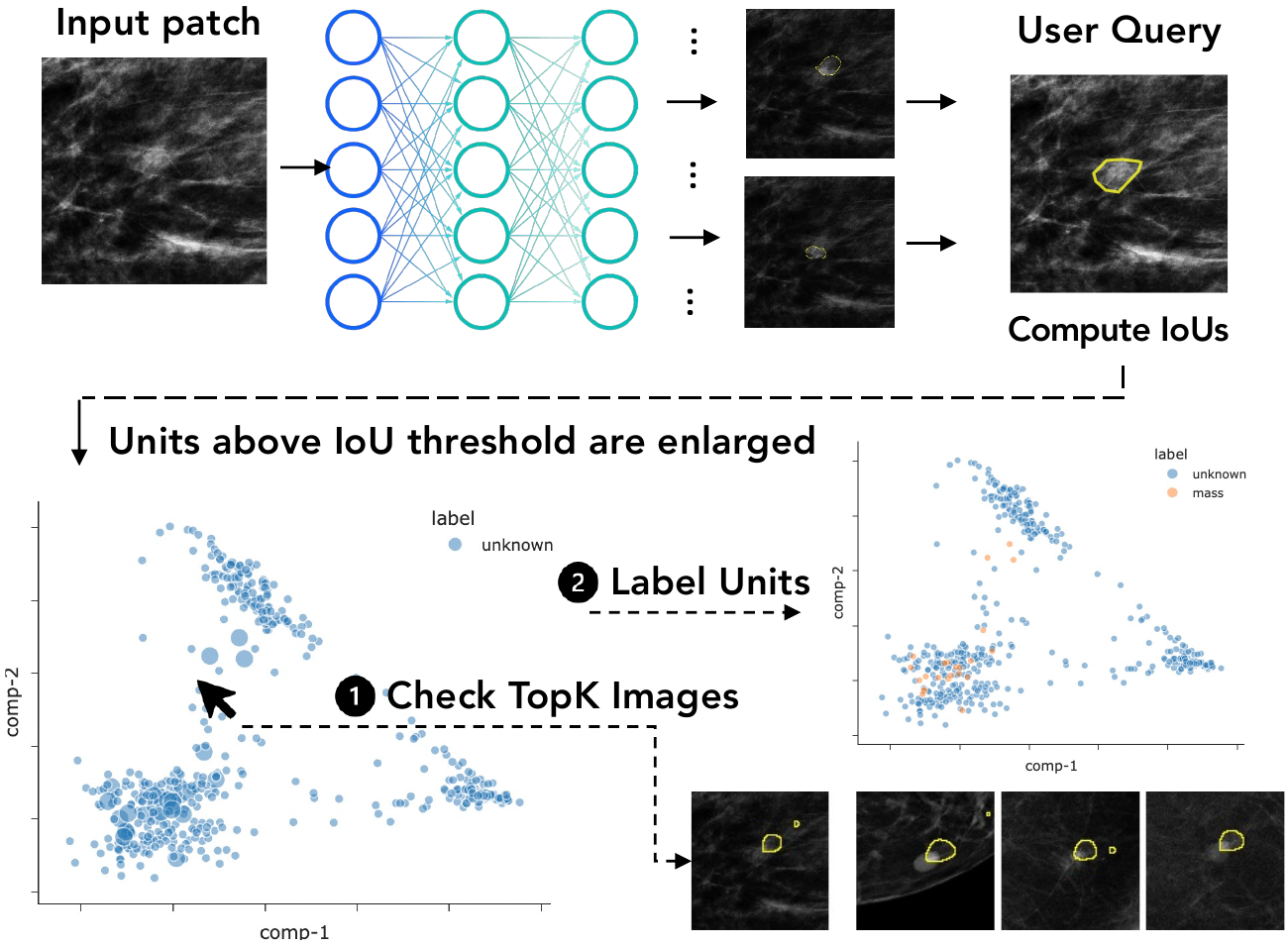}
    \caption{We provide an illustration of our system's main workflow. Each time a patch is feed to the model, its feature maps and corresponding activation maps in the last layer will be retained. When the user initiated a query, the user-defined activation map will be compared with all the retained activation maps based on IoU. Neurons whose activation maps have a high IoU (above the threshold) with one drew by the user will be highlighted in the scatter plot. The user can then click on these neurons to check their top activated images. When the user is convinced, they can annotate these neurons with the concept they detect. These annotations will be used in generating explainability reports in the interface.}
    \label{fig:proc}
\end{figure*}

\section{System Design}
The key challenge in deciding the semantic meaning of neurons in a CNN is their shear number. Classic CNNs often have tens or even hundreds of layers, each containing between 512 and 2048 neurons. While previous work \cite{bau2020understanding} has found that neurons in the last CNN layer tend to have the highest level of interpretability, it is still a label-intensive and time-consuming process for domain experts to inspect and label concepts for hundreds if not thousands of neurons in the last layer. To deal with these challenges, we propose an interactive system that takes advantage of the Network Dissection methodology for users to understand their model while avoiding the costly overhead of inspecting and labeling individual neurons. An overview of our interface is shown in Fig \ref{fig:teaser}, and the design goals behind our system are listed in the following sections.

\subsection{Goal 1: Query Neuron's Learned Representations}
\label{subsec:query}
The primary goal of our proposed system is to allow radiologists to answer the question: are there neurons that activate on features that users deem important? A key hypothesis to our query mechanism is that neurons with similar semantic meanings tend to have similar activation maps on a given image. Thus, the problem of finding neurons that align with human reasoning boils down to identifying neurons whose activation maps have a high degree of overlap with regions deemed important by a domain expert for a given image. 

\subsubsection{Activation Map}

After identifying top activated images for each neuron, we need to show an activation map for each image to signal what features in the image cause a neuron to activate. We followed the same neuron visualization technique as \cite{bau2020understanding}. Essentially, using the same notation from the previous section, we can compute a global activation quantile $q_{n_i}$ for each $n_{i}$ on all images in $X$. Then, for each $x_{i}$, we can generate the activation map by highlighting pixels with an activation value higher than a predetermined quantile value. We used the 99\% quantile value to generate activation map of images for each neuron.

\subsubsection{Query Metric} Given our proposed query mechanism, the metric we used to compare the salient regions defined by the user and the activation maps of each neuron is Intersection over Union (IoU), which is widely used in evaluating segmentation tasks. We use $A$ to denote the user-defined region and $B$ to denote the activation map of a neuron. Then, IoU can be defined as the following:

\begin{equation}
    IoU = \frac{\mid A \cap B \mid}{ \mid A \cup B \mid}
\end{equation}

\subsection{Goal 2: Label Meaningful Neurons for Explanation}
\label{subsec:neuron_annot}
\subsubsection{Label Groups of Neurons}
Another function that our interface provides is that it allows the user to label neurons. The process is as follows: after the user highlights a region in the image, we compute the IoU of the region and the activation maps of all neurons. If a neuron's activation map has an IoU score above our specified threshold (0.2), we increase the size of the dot of the corresponding neuron in the scatter plot. Then, the user will be able to click on these neurons to see their top activated images. If these neurons indeed detect consistent concepts, the user can label these neurons. If the concept is not yet provided in the dropdown menu, the user has the option to add a new one. 

\subsubsection{Generate Explainability Reports}
One of the key benefits of labeling semantically meaningful neurons is the promise of detailed explainability reports. We implemented two ways of generating explainability reports using neurons' labels to provide users with additional insights. To streamline the workflow, our strategy is purely post-hoc and thus is different from concept extraction strategies such as \cite{zhao2021human} that leverage active learning.  

With neurons and associated labels, we can compute the mean max activation values of neurons belonging to the same label. Given an image containing a medical phenomenon $P$, we should expect that the mean activation of neurons labeled as $P$ have higher values than those of neurons with other labels. We show this information with a bar chart, as in section (6) of Fig \ref{fig:teaser}, under ``activation value".

The other explainability report depends on the user-defined region $S$ of the incoming image. With this additional input, we can compute the mean IoU between $S$ and activation maps of neurons with the same label. Similarly, given an image containing medical phenomena $P$, we should expect the mean IoU between $S$ and activation maps of neurons labeled $P$ will have higher values than those of neurons with other labels. Similarly, we use a bar chart to encode this information, as in section (6) of Fig \ref{fig:teaser}, under ``activation area".

\subsection{Goal 3: Decode Semantic Connections of Neurons}
Finally, to help users understand the relation between neurons and view each neuron's highly activated images, we introduced an embedding for neurons and utilized a dimension reduction technique.

\label{subsec:neuron_vis}
\subsubsection{Neuron Embedding}
 Our method is based on the intuition that, on a diverse set of images, neurons detecting different concepts will have different activation maps on these images. Therefore, given a test set $X$ with $n$ samples, for each neuron, we can compute the maximum activation value on all $n$ samples. Each neuron will be associated with a discriminative vector in $\mathbb{R}^{n}$. If we have $m$ units, we end up with an embedding of shape $m \times n$. While previous work \cite{park2021neurocartography} also explored the idea of neuron embedding, our method presents a more computationally efficient solution without the need for an optimization process.

\subsubsection{Dimension Reduction for Visualization}
For visualization purposes, we apply PCA with 2 components, generating a matrix of shape $m\times 2$. The visualization is shown in the Label Units section of the interface. Each dot in the scatter plot represents a neuron in the last layer of our model.

\subsection{Implementation Details}
We implemented the system purely in Python. We used PyTorch \cite{paszke2019pytorch} to develop models and Plotly Dash \cite{plotly} to build the front-end.

\section{Interface Workflow}
In this section, we will provide an overview of how our interface can potentially fit into a radiologist's workflow (Fig. \ref{fig:teaser}). 

In the Mammogram component on the left, the radiologist can browse all the mammograms in the data set. Each time the radiologist switches to a new mammogram, our system's backend will split it into non-overlapping $512 \times 512$ patches and feed them to the trained model for inference. The lesion patches are shown in area (1). If the user wants to have a closer look at a specific patch, they can click on it in (1). Then, the corresponding region will be highlighted in the full mammogram to provide context as in (2).

Furthermore, an enlarged version of the selected patch will be presented in area (3), together with the softmax score. In addition, the activation map of the most activated neuron on the input image will appear in the middle of the Query Units component. With this information, the radiologist will get a sense of what features of the patch led to the model's final decision. 

At this stage, the radiologist may be satisfied with the justification provided by the activation map and can simply move on to subsequent mammograms. However, the activation map may not be perfect. As we discussed in Section 4, the user might be interested in knowing whether there are neurons in the model that focus on a region that is not well covered by the activation map of the most activated neuron. In this case, our system provides a solution by allowing the radiologist to define their region of interest in (3). Once the region is defined, our system will follow the methods laid out in section \ref{subsec:query} to identify neurons whose activation map has a high overlap with what the user defined. The most aligned activation map is then shown on the right of the Query Units component.

When the IoU scores between the user-defined region and all neurons' activation maps are computed, the scatter plot in (4) will be updated. As discussed in \ref{subsec:neuron_vis}, this scatter plot is a 2D projection of our proposed neuron embedding using PCA. This plot, with each point denoting a specific neuron, provides several important interactions to assist users' visual exploration of the model's learned representations (Fig \ref{fig:proc}). When the user selects the patch for further investigation, the patch will be fed to the model, and an activation map will generated for each of the neuron in the last convolutional layer. If the user selects a salient region and queries neurons with similar semantic meanings, the system backend will compute the IoU between the region and activation maps of all neurons. The points representing neurons whose activation maps have high IoU values (over the specified threshold) will be enlarged. Then, the user may click on the relevant point to inspect both the neuron's activation map on the input patch and its top activated images. In this way, the user can confirm whether a neuron consistently captures a concept. When the user is convinced, they may endow those neurons with a label for the concept they detect, after which the points will return to their original size but put on a different color. The user can perform the annotation in (5), where they can either select preexisting labels in the drop-down menu or add new labels if needed. As the user can label these units in groups, little annotation effort is required. 

While the above workflow allows the user to gain understanding of the model's decision-making process, the neuron annotations can be used to generate explainability reports in (6) as discussed in section \ref{subsec:neuron_annot}. Notably, the user does not need to label all neurons to generate such explainability reports; instead, they can label neurons gradually as they perform diagnosis to help the system generate better, finer-grained reports over time. 

\section{Conclusion}
In this paper, we proposed a novel visual interface to leverage state-of-the-art interpretability techniques to help radiologists screen for breast cancer in an AI-assisted workflow. Instead of letting experts passively interpret saliency maps from a deep learning model, our system allows them to actively query relevant learned representations to understand the model's decisions. With the proposed neuron embedding and visualizations, users can easily inspect the model's learned representations and provide annotations to groups of neurons with minimal effort. We demonstrate that our system can leverage user annotations to provide explainability reports naturally as they perform diagnosis over time. We believe that our system sheds light on how we can leverage the fruitful research in machine learning to maximize its potential in transforming healthcare by human-centric software that considers its applications in actual clinical settings. Meanwhile, we point out that our system currently lacks extensive evaluations. On the user side, it is critical to evaluate whether they indeed feel more transparency about the model's decision with various informative interactions provided by the interface. In terms of system design, an interesting alternative to labeling neurons is to train a finer-grained classification model with the same patches in an active learning setting, which might also produce fine-grained reports with little annotation effort. We intend to pursue these ideas in our future work.

\acknowledgments{This work used the Extreme Science and Engineering Discovery Environment (XSEDE) \cite{towns2014xsede}, which is supported by National Science Foundation grant number ACI-1548562. Specifically, it used the Bridges-2 system, which is supported by NSF award number ACI-1445606, at the Pittsburgh Supercomputing Center (PSC).}

\bibliographystyle{abbrv-doi}

\bibliography{main}
\end{document}